\documentclass[preprint,12pt]{elsarticle}
\usepackage{amssymb}
\usepackage{amsmath}
\usepackage{graphicx}% Include figure files
\usepackage{dcolumn}% Align table columns on decimal point
\usepackage{bm}% bold math
\usepackage{caption}
\usepackage{subcaption}

\usepackage{float}\usepackage{xcolor}
\usepackage{booktabs}\usepackage{multicol}
\journal{Zeitschrift f\"ur Naturforschung A}

\begin{document}

\begin{frontmatter}

\title{Investigation of $^{90, 92}$Zr(n,$\gamma$)$^{91, 93}$Zr in the $s$-process nucleosynthesis.}
\author{{Abdul Kabir$^{1}$, Zain Ul Abideen$^{1}$, Jameel-Un Nabi$^{2}$, and Dawar Khan$^{1}$}}

\address{$^{1}${Space and Astrophysics Research lab, National Centre of GIS and Space Applications, Department of Space Science, Institute of Space Technology, Islamabad 44000, Pakistan}}

\address{$^{2}${University of Wah, Quaid Avenue, Wah Cantt 47040, Punjab, Pakistan}}

                    %%%%%%%%%%%%   ABSTRACTT %%%%%%%%%%%%%%
\begin{abstract}
{The neutron capture rates and Maxwellian-averaged cross-sections (MACS) for $^{90}$Zr(n,$\gamma$)$^{91}$Zr and $^{92}$Zr(n,$\gamma$)$^{93}$Zr processes have been computed within the framework of Talys v1.96. The effects of phenomenological nuclear level density (NLD) parameters and the gamma strength functions (GSFs) on Maxwellian-averaged cross-sections and neutron capture rates {are} examined both quantitatively and qualitatively. The present model-based computed data for MACS and reaction rates gives a good comparison with the existing literature. 
{The fine-tuning of the statistical model's nuclear properties (level density and gamma-ray strength) to reproduce experimental data {will allow the detailed investigation of the $s$- process network}.}}
\end{abstract}
\begin{keyword}
{Cross-section, AGB stars, MACS, talys, nuclear level density, nuclear rates}
\end{keyword}
\end{frontmatter}
%%%%%%%%%%%%%%%%%%%%%%%%%%%%%%INTRODUCTION%%%%%%%%%%%%%%%%%%%%%%%%%%%%%%%%%%%%%%

\section{Introduction}
Nuclei more massive than iron are mostly formed by neutron capture reactions, which are called after their widely distinct time scales as the $s$-process (slow neutron capture process) and the $r$-process (rapid neutron capture process) \cite{Burbidge,Cameron}.  {The $s$-process in the present model of stellar nucleosynthesis is a bit more challenging than previously believed~\cite{Kappeler11}. The elements from Fe to Zr (A$\leq$90) are {formed} by the so called weak component, which is driven by neutrons produced through the $^{22}$Ne($\alpha$,n)$^{25}$Mg reaction in massive stars. {Thus, a stellar model to describe the weak $s$-process requires accurate and reliable {MACS} at $s$- process relevant temperatures}. The main $s$-process operates in the AGB stars and produces the heavier isotopes up to the lead-bismuth {region}. Because of the breakthroughs in astronomical observations and stellar modeling, accurate assessments of the $s$-process have become incredibly influential~\cite{Lugaro}. The current stellar model suggests that the neutron production for the weak $s$-process occurs in two different burning stages, which are the He core burning and the Carbon shell burning~\cite{Raiteri}. The burning temperatures of the two stages are not the same (T$_9$=0.3 for the He core burning and T$_9$=1 for the carbon shell burning). Hence, the $s$-process takes place when $0.3<\rm T_9<1$ and at neutron density {of} $N_n$$\approx$ $10^8$~cm$^{-3}$~\cite{Kappeler90}. 
%	The weak $s$-process is believed to take place in massive stars and produces most of the $s$-abundances in the mass region between Fe and Zr. {Thus, a stellar model to describe the weak $s$-process requires} accurate and reliable {MACS}, not only at temperature  T$_9$=0.3 (kT=25 keV) but also at T$_9$=1 (kT=90 keV). The main $s$-process operates in the AGB stars and produces the heavier $s$-process isotopes up to the lead-bismuth. Because of breakthroughs in astronomical observations and stellar modeling, accurate assessments of the $s$-process have become incredibly influential~\cite{Lugaro}. 
The $s$-process has evolved from a simple explanation of the abundance distribution of elements in the solar system to a more comprehensive {description} that takes into account the general features of stellar and galactic dynamics. As a result of such advancements, the $s$-process has become a powerful tool for examining the evolution of red giant stars~\cite{Busso}.}

Here, we focus the reader's attention towards the neutron capture cross-sections {of} two isotopes of zirconium.  Substantial progress has been achieved in the analysis of Zr from the  astrophysical perspectives. Advances in nuclear modeling and observational techniques have allowed researchers to gain a better understanding of the nucleosynthesis of heavy elements in stars. Zr isotopes mostly have the $s$-process origin and they are associated with the first $s$-process peak in the solar abundance distribution around $A$$\approx$90. Both $^{90}$Zr and $^{92}$Zr are {key to the} formation of heavy isotopes. 
%$s$-process occurs in the late stages of stellar evolution when stars have exhausted their primary fuel and have begun {fusing} heavier elements. Understanding the evolution of stars and the origin of heavy elements in the universe is thus dependent on the study of these nuclei. 
%The stellar neutron capture rate is directly influenced by the neutron capture cross-section $\sigma(E_n)$~\cite{Akira}.
The neutrons are quickly thermalized in the He-burning stellar plasma, where the $s$-process is carried out. Nevertheless, the reaction flow is constrained by the extremely small neutron capture cross-sections of the neutron magic nuclei. $^{90}$Zr and $^{92}$Zr share the magic number and nearly magic number neutrons $N$=50, 52. The mass area of about $N$=50 is particularly fascinating in an astrophysical scenario~\cite{Koloczek}. Thus, their cross-sections contribute to the initial bottleneck in the reaction flow from the Fe seed to heavier isotopes. Tagliente \textit{et al.}~\cite{Tagliente1} examined the MACS of $^{90}$Zr(n,$\gamma$)$^{91}$Zr at $s$-process temperature by folding the capture cross-section with the thermal stellar spectra across a sufficiently broad neutron energy range (0.1-500)~keV, which is the highest temperature achieved during carbon shell burning in massive stars. The same measurement has been performed by authors in Ref.~\cite{BOLDEMAN1} within (3-200)~keV. 
Similarly, it could be argued that $^{92}$Zr(n,$\gamma$)$^{93}$Zr also has an $s$-process origin. It was challenging to verify the MACS values for this process. The majority of the experimental data currently available originate from measurements that are rather old. The authors in Refs.~\cite{OHGAMA,GOOD,Bartolome} measured the cross-section for the Zr nuclei including $^{92}$Zr. Similarly, the authors in Ref.~\cite{Tagliente} have measured the cross-section for $^{92}$Zr(n,$\gamma$)$^{93}$Zr over a broad range of neutron energies by utilizing the innovative features of the $\rm n\_TOF$ facility at CERN. The cross-section determined in their investigation leads to a reduction in the $s$-component of $^{92}$Zr. 

{In the present study, both the $^{90}$Zr(n,$\gamma$)$^{91}$Zr and $^{92}$Zr(n,$\gamma$)$^{93}$Zr reactions have been analyzed within the framework of the statistical model code Talys v1.96~\cite{TALYS1.96}. The Talys v1.96 code is based on the Hauser-Feshbach theory~\cite{hf1, hf2}. The main inputs in the Hauser-Feshbach theory are the NLDs, the optical model potentials (OMPs), and the gamma strength functions (GSFs). For low-energy neutrons as incident particles, the effect of changing the OMPs can be ignored in favor of the other two components (NLDs and GSFs) \cite{hfsr}. The optical model employed in this study is the local OMP by Koning and Delaroche \cite{kd}. The results were evaluated for all combinations of the {phenomenological} NLDs (constant temperature model (CTM), back-shifted Fermi gas model (BSFM), generalized superfluid model (GSM) \cite{ctm, bfm, gsm1, gsm2}) and GSFs ({Kopecky-Uhl Lorentzian, Brink-Axel Lorentzian,} and Gogny D1M QRPA \cite{brink1, brink2, kop, gogny}).}
{The predictions of the Talys v1.96 {code} have been compared with the evaluated nuclear data libraries including ENDF, TENDL, and JENDL, as well as the existing experimental data. The MACS have been computed with neutron energies up to 100 keV. The abundances for the $s$- and the $r$- processes may be reproduced through the sensitivity simulations of these processes. However, one of the major sources of uncertainty lies in the correct estimation of nuclear properties such as the neutron capture cross-section and rates for the nuclei involved in these processes.}
%%%%%%%%%%%%%%%%%%%%%%%%  %%%%%%%%%%%%%%%%%%%%%%%%%%%%%%%%%
\section{Theoretical Framework}
{The Talys v1.96 {code} used for the simulation of nuclear reactions, includes several state of the art nuclear models to cover almost all key reaction mechanisms encountered for light particle-induced nuclear reactions. It provides an extensive range of {reaction channels}. The possible incident particles {can be simulated} in the {$E_i$=(0.001–200)~MeV energy range}, and  the target nuclides {can be} from $A$=12 {onwards}. The output of the nuclear reaction includes fractional and total cross-sections, angular distributions, energy spectra, double-differential spectra, MACS, and capture rates.  Radiative capture is important in the context of nuclear astrophysics in which a projectile {fuses} with the target nucleus and {emits $\gamma$- ray} \cite{Kabir1,Kabir2,Kabir3,Kabir4,Kabir5}. The nuclear cross-section is an important factor in the calculation of radiative capture rates. The Maxwellian averaged cross-section is used when the energies of the projectiles follow a Maxwellian distribution, like in the stellar environment. MACS is an average of the cross-section over a range of energies, weighted by the Maxwell-Boltzmann distribution.}
\begin{align}
	{\langle \sigma \rangle (kT) = \frac{2}{\sqrt{\pi}(kT)^2}\int_0^\infty E \sigma (E) exp(\frac{-E}{kT}) dE.}
\end{align}
{where $k$, is the Boltzmann constant, $T$ is the temperature, $\sigma(E)$ is the capture cross-section and $E$ is the projectile energy.}
{In statistical models for predicting nuclear reactions, level densities are needed at excitation energies where discrete level information is not available or {is} incomplete. Together with the optical model potential, a correct level density is perhaps the most crucial ingredient for a reliable theoretical analysis of cross-sections, angular distributions, and other nuclear {quantities}. Six different level density models, among them three phenomenological and the {rest microscopic}, are {included in the Talys code}. For each of the phenomenological models, the constant temperature model (CTM), the back-shifted Fermi (BSFM) gas model, and the generalized superfluid model (GSM), a version without explicit collective enhancement is considered. In the CTM, the excitation energy range is divided into low energy regions i.e., {from 0 keV up to the matching energy ($E_{M}$) and high energy above the $E_{M}$ where the fermi-gas model (FGM) applies}. Accordingly, the constant temperature part of the total level density reads as.} 
\begin{align}
	{\rho_{T}^{tot}(E_x) = \frac{1}{T}exp(\frac{E_x-E_o}{T})}
\end{align}
{where $T$ and $E_o$ serve as adjustable parameters in the constant temperature expression.  The BSFM is used for the whole energy range by treating the pairing energy as an adjustable parameter.}
\begin{align}
{\rho_{F}^{tot}(E_x) = \frac{1}{\sqrt{2\pi}\sigma}\frac{\sqrt{\pi}}{12}\frac{exp(2\sqrt{aU})}{a^{1/4}U^{5/4}}}
\end{align}
{where $\sigma$ is the spin cut-off parameter, which represents the width of the angular momentum distribution, $U$ is the  effective excitation energy and $a$ is the level density parameter defined below.} 
\begin{align}
	{a = \tilde{a}(1+\delta W \frac{1-exp(-\gamma U)}{U})}
\end{align}
{where $\tilde{a}$ is the asymptotic level density without any shell effects. $\delta W$ gives the shell correction energy, and the damping parameter $\gamma$ determines how rapidly $a$ approaches to $\tilde{a}$. One should note that for the best fitting one can readjust the $a$ to achieve the desired value of cross-section and nuclear reaction rates. For further investigations, one can {refer to}~\cite{koni}. The GSM is similar to CTM in the way that it also divides the energy range into low and high energies. The high energy range is described by BSFM as mentioned before. It is characterized by a phase transition from a superfluid behavior at low energies, where the pairing correlations strongly influence the level density.} 
\begin{align}
{	\rho^{tot}(E_x) = \frac{1}{\sqrt{2\pi}\sigma}\frac{e^S}{\sqrt{D}}}
\end{align}
{where $S$ is the entropy and $D$ is the determinant related to the saddle-point approximation.}

{The gamma strength function (GSF) plays a crucial role in estimating cross-sections and reaction rates, particularly in processes involving the emission of gamma rays.  Different GSFs are included in Talys v1.96, among them the Brink-Axel model is used for all transitions except for $E1$ \cite{TALYS1.96}. The GSF function $f_{XL}$ gives the distribution of the average reduced partial transition width as a function of the photon energy $E_{\gamma}$ \cite{RSF2}. }
\begin{align}
	{f_{XL}(E_{\gamma}) = K_{XL}\frac{\sigma_{XL}E_{\gamma}\Gamma^2_{XL}}{(E_{\gamma}^2-E_{XL}^2)^2+(E_{\gamma}\Gamma^2_{XL})^2}}
\end{align}
{where $E_{XL}$ is the evergy, $\Gamma_{XL}$ is the width, and $\sigma_{XL}$ is the giant resonance strength. For $E1$ transitions, Talys v1.96 utilizes the {Kopecky-Uhl} model by default.}  
\begin{align}
	f_{XL}(E_{\gamma}, T) &= K_{XL}\big[\frac{E_{\gamma}\tilde{\Gamma}_{E1}(E_{\gamma})}{(E_{\gamma}^2-E_{E1}^2)^2+E^2_{\gamma}\tilde{\Gamma}_{E1}(E_{\gamma})^2}+\frac{0.7\Gamma_{E1}4\pi^2T^2}{E_{E1}^3}\big]\sigma_{E1}\Gamma_{E1}
\end{align}
and
\begin{align}
	\tilde{\Gamma}_{E1}(E_{\gamma}) &= \Gamma_{E1}\frac{E_{\gamma}^2+4\pi^2}{E_{E1}^2}\big[\frac{E_n+S_n-\Delta-E_{\gamma}}{a(S_n)}\big]
\end{align}
{where $\tilde{\Gamma}(E_{\gamma})$ represents the energy-dependent damping width, $E_n$ is the incident energy of neutrons, $S_n$ is the neutron seperation energy, and $\Delta$ is the correction for pairing.  $a$ represents the level density parameter at $S_n$ as mentioned earlier.}

\section{Results and Discussion} 
To investigate the impact of $^{90}$Zr(n,$\gamma$)$^{91}$Zr and $^{92}$Zr(n,$\gamma$)$^{93}$Zr reactions in the $s$-process of nucleosynthesis, we have computed the MACS and neutron capture rates via different nuclear level density models and gamma strength functions. Zr isotopes have almost magic neutron configurations, which results in their small (n,$\gamma$) cross-sections and fairly large $s$-abundances. We employed the Talys v1.96 for the analysis of MACS. Talys v1.96 enabled us to evaluate nuclear reactions from the unresolved resonance range up to intermediate energies. The MACS of $^{90}$Zr(n,$\gamma$)$^{91}$Zr and $^{92}$Zr(n,$\gamma$)$^{93}$Zr were computed via three different GSFs and NLDs models available in Talys v1.96. Theoretical MACS can be computed with Talys in the Hauser-Feshbach framework by choosing the NLD, and a radiative strength function. Out of several possibilities for level density calculations, the CTM, BSFM, and the GSM for excitation energies up to 100~keV were employed in our analysis. The obtained NLDs computed by the CTM, BSFM, and GSF were employed for the calculation of MACS and capture rates of  $^{90}$Zr(n,$\gamma$)$^{91}$Zr and $^{92}$Zr(n,$\gamma$)$^{93}$Zr. One should note that for each NLDs, we have fixed the  GSFs. For  example, we have fixed the GSFs  as Brink-Axel, Gogny, and  {Kopecky-Uhl},  and changed the NLDs by the CTM, BSFM, and GSM,  respectively.  
%%%%%%% FIG. (1)       %%%%%%%%%%%%%%%%%
\begin{figure}
\centering
\includegraphics[width=1.4\textwidth]{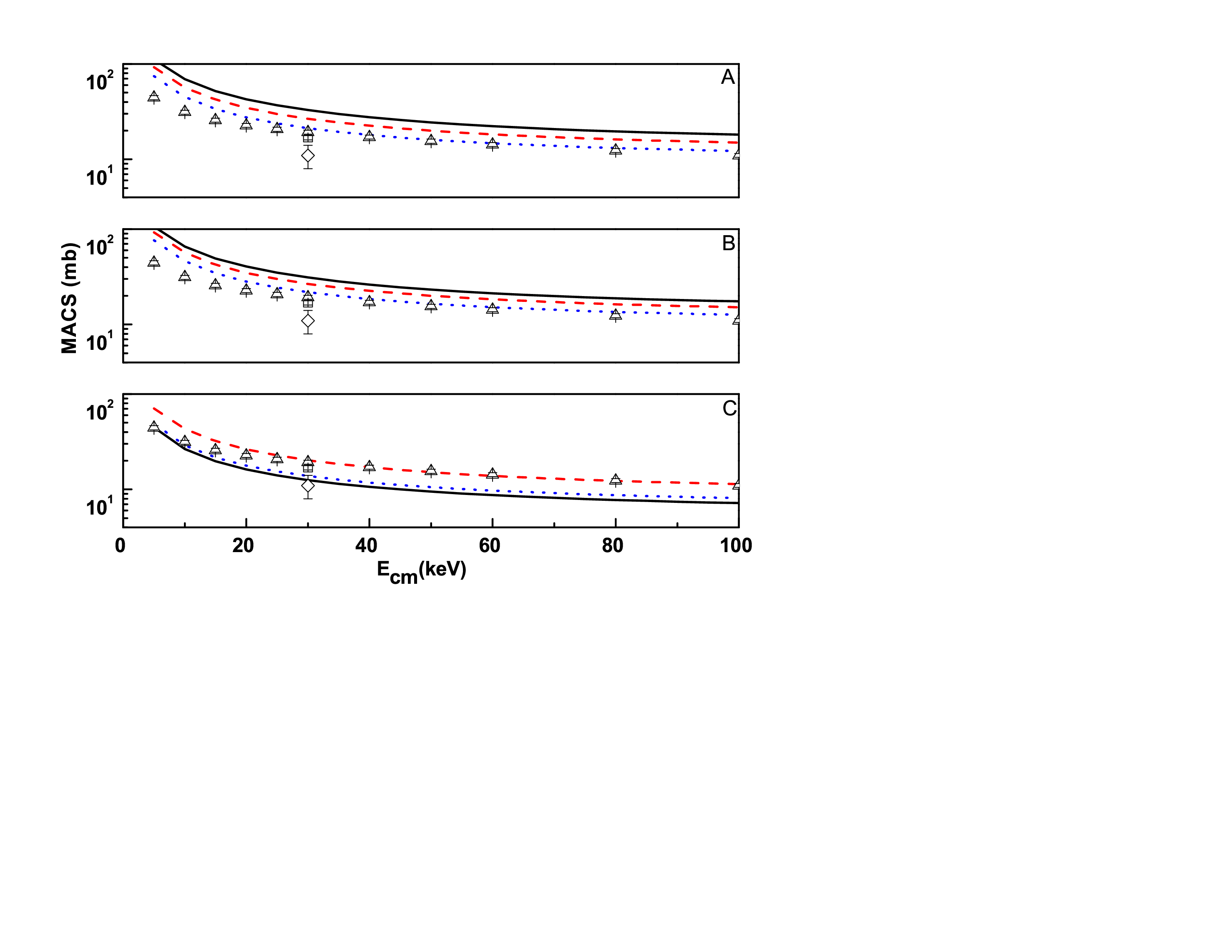}
\vspace*{-60mm}
\caption{{{{The total MACS for $^{90}$Zr(n,$\gamma$)$^{91}$Zr along with the measured data \cite{Tagliente1}($\bigtriangleup$), \cite{BOLDEMAN1} ($\square$) and \cite{Macklin} ($\diamond$). (a) Computed data by the CTM with the GSFs as Gogny (solid line), {Brink-Axel (dashed line)  }and {Kopecky-Uhl} (dotted line). (b) Computed data by the BGSM with the GSFs as {Gogny (solid line),  Brink-Axel (dashed line)} and  {Kopecky-Uhl} (dotted line). (c) Computed data by the GSM with the GSFs as Gogny (solid line),  Brink-Axel {(dashed line)} and  {Kopecky-Uhl} (dotted line).}}}}
\label{fig:1}
\end{figure}

\begin{figure}
	\centering
	\includegraphics[width=1.4 \textwidth]{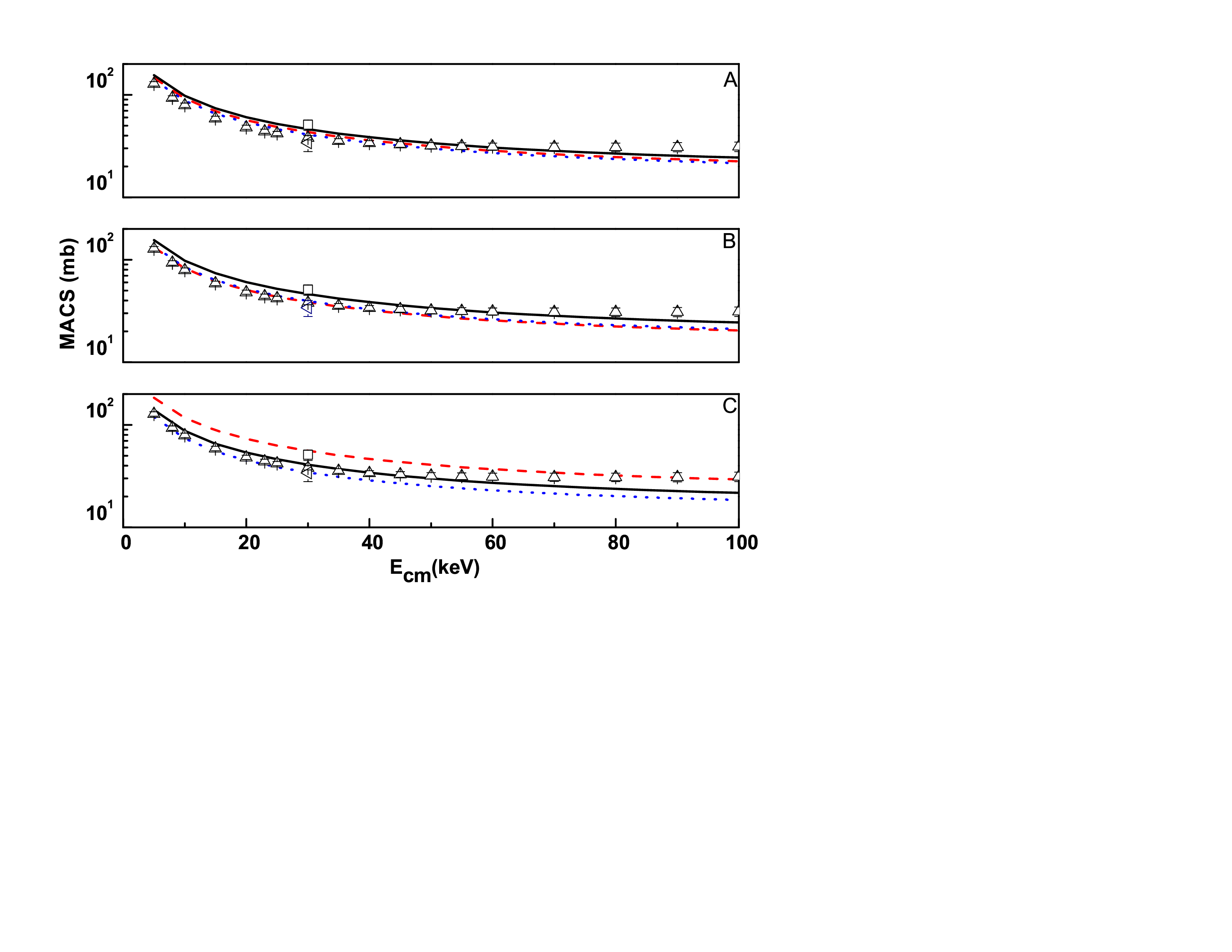}
	\vspace*{-60mm}
	\caption{{{{The total MACS for $^{92}$Zr(n,$\gamma$)$^{93}$Zr along with the experimental data Refs.~\cite{BOLDEMAN1} ($\square$), \cite{Tagliente} ($\bigtriangleup$),  and \cite{Macklin} ($\diamond$). (a) Computed data by the CTM with the GSFs as Gogny (solid line), {Brink-Axel (dashed line)  }and {Kopecky-Uhl} (dotted line). (b) Computed data by the BGSM with the GSFs as {Gogny (solid line),  Brink-Axel (dashed line)} and {Kopecky-Uhl} (dotted line). (c) Computed data by the GSM with the GSFs as Gogny (solid line),  Brink-Axel {(dashed line)} and  {Kopecky-Uhl} (dotted line).}}}\label{fig:2}}
\end{figure}

%%%%%%%%%%%%%%%%%%%%%%%%%%% FIG. (2) %%%%%%%%%%%%%%%%%%%%%
{ In the $^{90}$Zr(n,$\gamma$)$^{91}$Zr analysis, we have fixed the GSF as {Brink-Axel}, and varied the NLDs as CTM, BSFM and GSM. Their results are mentioned in Figs.~(\ref{fig:1})a-c. The computed MACS at $kT$=30 keV were 26.69~mb (CTM), 26.75~mb (BSFM), and 12.54~mb (GSM). For the GSFs as Gogny, the computed MACS at $kT$=30 keV were 32.83~mb (CTM), 31.15~mb (BSFM), and 20.29~mb (GSM). Similarly, for the GSF as {Kopecky-Uhl}, the computed MACS at $kT$=30 keV were 21.29~mb (CTM), 21.84~mb (BSFM), and 13.83~mb (GSM).  The present analysis are within the range of the $20.8\pm2.1$ \cite{compiled}, $21\pm2$~mb \cite{Bao}, and more recent results $19.3\pm0.9$~mb \cite{Tagliente1}.  The MACS computed via different GSFs and phenomenological {model} based NLDs along with the experimental data are depicted in Figs.~(\ref{fig:1}). Fig.~(\ref{fig:1})a, depicts the computed MACS via CTM and GSFs as  Gogny (solid line), {Brink-Axel} {(dashed line)}, and {Kopecky-Uhl} (dotted line). Fig.~(\ref{fig:1})b, depicts the computed MACS via BSFM and GSF as  Gogny (solid line), {Brink-Axel} {(dashed line)}, and {Kopecky-Uhl} (dotted line). Similarly, Fig.~(\ref{fig:1})c, depicts the computed MACS via GSM and GSFs as  Gogny (solid line), {Brink-Axel} {(dashed line)}, and {Kopecky-Uhl} (dotted line). {Amongst all three cases,} the {Kopecky-Uhl} Lorentzian model of Talys is the best fit GSF model for {the} $^{90}$Zr(n,$\gamma$)$^{91}$Zr process.} Furhermore, we have accounted {for} the percentage difference {in} our analysis and {those of} \cite{Tagliente1} at $kT$=30 keV. It was noted that when the {Kopecky-Uhl} Lorentzian model (NLDs {as} CTM and BSFM) of Talys code is employed for $^{90}$Zr(n,$\gamma$)$^{91}$Zr process, the percentage difference between our analysis and those of \cite{Tagliente1} at $kT$=30 keV are 3.5\% (GSF as {Kopecky-Uhl} Lorentzian model and NLD {as CTM}) and 7.5\% (GSF as {Kopecky-Uhl} Lorentzian model and NLD as BSFM).
%%%%%%%%%%%%%%%%%%%%%%%%%%%%%%%%%%%%%%%%

{Similarly, the same sets of GSFs and NLDs (as mentioned above) were employed for the analysis of $^{92}$Zr(n,$\gamma$)$^{93}$Zr process. Fixing the GSF as {Brink-Axel},  the computed MACS at $kT$=30 keV were 43.22~mb (CTM), 38.52~mb (BSFM), and 55.86~mb (GSM). For the GSF as Gogny, the computed MACS at $kT$=30 keV were 49.70~mb (CTM), 46.21~mb (BSFM), and 40.95~mb (GSM). Similarly, for the GSF as Kopecky, the computed MACS at $kT$=30 keV were 40.90~mb (CTM), 39.44~mb (BSFM) and 34.32~mb (GSM). The  present analysis for $^{92}$Zr(n,$\gamma$)$^{93}$Zr {is} within the range of $33\pm4$~mb \cite{Bao}, {$46\pm5$~mb \cite{Nakagawa}} and the most {recent measurements} $38\pm3$ \cite{Tagliente}. Further comparisons are illustrated in Figs.~(\ref{fig:2})a-c along with the measured data.} {We have accounted for the percentage difference between our {analysis} and those of \cite{Tagliente1} at $kT$=30 keV. It was noted that when the {{Kopecky-Uhl}} Lorentzian model (NLDs {as} CTM and BSFM) of Talys is employed for $^{92}$Zr(n,$\gamma$)$^{93}$Zr, the percentage difference between our analysis and those of \cite{Tagliente} at $kT$=30 keV were 16.3\% ({GSF as}  {{Kopecky-Uhl}} Lorentzian model and NLD {as} CTM) and 14.7\% ({GSF as}  {{Kopecky-Uhl}} Lorentzian model and NLD {as} BSFM).}

\begin{table}
\centering
\caption{Comparison between the present best-fit analysis and those of standard codes or evaluated data Ref.~\cite{RSF2} for $^{90}$Zr(n,$\gamma$)$^{91}$Zr and $^{92}$Zr(n,$\gamma$)$^{93}$Zr processes. }
\label{tab:ENDF1}       
\addtolength{\tabcolsep}{1pt}
\scalebox{0.5}{
{\begin{tabular}{cccccccccc}
\toprule
\multicolumn{1}{c}{} & \multicolumn{1}{c}{} & \multicolumn{8}{c}{{$^{90}$Zr(n,$\gamma$)$^{91}$Zr}} \\
\cmidrule(rl){2-10} 
{$k_BT$} & This work& ENDF/B-VII.0 & ENDF/B-VII.1 &JEFF-3.1 & JENDL-3.3 & JENDL-4.0&ENDF/B-VI.8&ROSFOND& CENDL-3.1	
\\
& &USA06 & USA11 &  Europe05 &  Japan02 & Japan10& USA01&RUSSIA08& China09 \\
\midrule
20&	27.57&	25.07&	23.71&	25.05&	25.05&	22.70&	26.43&	24.85&	25.05\\
25&	23.83&	22.03&	21.01&	22.64&	22.64&	20.69&	23.80&	22.58&	22.62\\
30&	21.29&	19.65&	18.93&	20.82&	20.82&	19.21&	21.97&	20.82&	20.79\\
35&	19.44&	17.78&	17.29&	19.38&	19.38&	18.04&	20.67&	19.40&	19.32\\
40&	18.04&	16.29&	15.97&	18.20&	18.20&	17.07&	19.71&	18.22&	18.09\\
45&	16.94&	15.09&	14.90&	17.19&	17.19&	16.25&	18.99&	17.21&	17.05\\
50&	16.06&	14.12&	14.02&	16.33&	16.33&	15.54&	18.42&	16.32&	16.14\\
55&	15.34&	13.32&	13.28&	15.57&	15.57&	14.92&	17.96&	15.53&	15.34\\
60&	14.75&	12.65&	12.67&	14.90&	14.90&	14.36&	17.58&	14.83&	14.63\\
65&	14.25&	12.08&	12.14&	14.30&	14.30&	13.87&	17.27&	14.20&	13.99\\
70&	13.83&	11.60&	11.69&	13.76&	13.76&	13.42&	16.99&	13.63&	13.42\\
75&	13.46&	11.19&	11.30&	13.27&	13.27&	13.02&	16.75&	13.11&	12.90\\
80&	13.15&	10.83&	10.96&	12.83&	12.83&	12.66&	16.54&	12.64&	12.44\\
85&	12.87&	10.52&	10.66&	12.42&	12.42&	12.34&	16.35&	12.21&	12.01\\
90&	12.64&	10.25&	10.40&	12.05&	12.05&	12.04&	16.19&	11.81&	11.63\\
95&	12.43&	10.01&	10.17&	11.70&	11.70&	11.77&	16.04&	11.45&	11.27\\
100&12.24 &	09.80&	09.96&	11.39&	11.39&	11.52&	15.90&	11.12&	10.95\\		
\end{tabular}}}
\addtolength{\tabcolsep}{1pt}
\scalebox{0.5}{
{\begin{tabular}{cccccccccc}
\toprule
\multicolumn{1}{c}{} & \multicolumn{1}{c}{} & \multicolumn{8}{c}{{$^{92}$Zr(n,$\gamma$)$^{93}$Zr}} \\
\cmidrule(rl){2-10} 
{$k_BT$} & This work& ENDF/B-VII.0 & ENDF/B-VII.1 &JEFF-3.1 & JENDL-3.3 & JENDL-4.0&ENDF/B-VI.8&ROSFOND& CENDL-3.1			
\\
& &USA06 & USA11 &  Europe05 &  Japan02 & Japan10& USA01&RUSSIA08& China09 \\
\midrule
20&	44.96&	57.67&	57.62&	57.67&	57.67&	55.45&	58.09&	61.56&	57.66\\
25&	38.66&	50.32&	50.19&	50.32&	50.32&	46.56&	48.79&	53.41&	50.27\\
30&	34.32&	45.72&	45.48&	45.72&	45.72&	40.49&	42.37&	47.88&	45.60\\
35&	31.14&	42.62&	42.26&	42.62&	42.62&	36.09&	37.72&	43.90&	42.42\\
40&	28.72&	40.41&	39.93&	40.41&	40.41&	32.76&	34.21&	40.91&	40.11\\
45&	26.82&	38.77&	38.16&	38.77&	38.77&	30.17&	31.48&	38.60&	38.36\\
50&	25.29&	37.50&	36.79&	37.50&	37.50&	28.10&	29.30&	36.75&	36.99\\
55&	24.03&	36.49&	35.68&	36.49&	36.49&	26.41&	27.53&	35.25&	35.87\\
60&	22.99&	35.66&	34.77&	35.66&	35.66&	25.01&	26.07&	34.00&	34.95\\
65&	22.11&	34.97&	34.01&	34.97&	34.97&	23.84&	24.84&	32.95&	34.18\\
70&	21.36&	34.38&	33.36&	34.38&	34.38&	22.84&	23.79&	32.05&	33.52\\
75&	20.72&	33.86&	32.81&	33.86&	33.86&	21.99&	22.89&	31.28&	32.96\\
80&	20.16&	33.41&	32.34&	33.41&	33.41&	21.25&	22.11&	30.60&	32.47\\
85&	19.68&	33.01&	31.92&	33.01&	33.01&	20.60&	21.43&	30.00&	32.04\\
90&	19.25&	32.65&	31.56&	32.65&	32.65&	20.03&	20.84&	29.48&	31.66\\
95&	18.87&	32.33&	31.24&	32.33&	32.33&	19.53&	20.31&	29.01&	31.32\\
100&	18.53&	32.03&	30.95&	32.03&	32.03&	19.09&	19.84&	28.59&	31.03\\			
\hline\hline
\end{tabular}}}
\end{table}

{Furthermore, we compared the best-fitted MACS with those of standard codes/evaluated data Ref.~\cite{RSF2} for $^{90}$Zr(n,$\gamma$)$^{91}$Zr and $^{92}$Zr(n,$\gamma$)$^{93}$Zr processes in Table.~(\ref{tab:ENDF1}). It was found that the present model-based results for the MACS are within the range of Ref.~\cite{RSF2}.} 
{{We have {also} shown the effect of the level density parameter ($a$) on the MACS. It was found that the adjustments had an effect on the MACS for both $^{90}$Zr(n,$\gamma$)$^{91}$Zr and $^{92}$Zr(n,$\gamma$)$^{93}$Zr. Lowering the adjustment parameter \textit{a} decreased the MACS for both, and vice versa. The change was also noticable at both low and high energies. These adjustments can be used to achieve perfect agreement between theory and experiment.}} 
%%%%%%%%%%%%%%%%%%%%%%                  %%%%%%%%%%%%%%%%%%%%%%%%%%%%%

\begin{table}[hbtp]
	\centering
	\caption{Effect of the adjustment parameter ($a$) on MACS for the $^{90}$Zr(n,$\gamma$)$^{91}$Zr and $^{92}$Zr(n,$\gamma$)$^{93}$Zr radiative capture processes. }
	\label{tab:CTM1}       
	\addtolength{\tabcolsep}{1pt}
	\scalebox{0.70}{
		{\begin{tabular}{ccccccc}
				\toprule
				\multicolumn{1}{c}{$^{90}$Zr(n,$\gamma$)$^{91}$Zr} 
				\\
				\cmidrule(rl){2-7}
				\multicolumn{1}{c}{} & \multicolumn{2}{c}{{BSFM}} & \multicolumn{2}{c}{{CTM}} &\multicolumn{2}{c}{{GSM}}\\
				\cmidrule(rl){2-3} \cmidrule(rl){4-5} \cmidrule(rl){6-7}
				$a$ & $k$T (MeV)& MACS (mb) & $k$T (MeV) & MACS (mb) & $k$T (MeV) & MACS (mb) \\
				\midrule
				0.5& 0.15 &   $2.00 \times 10^{-1}$  &  0.15 &  $1.92 \times 10^{-1}$ & 0.15 &  $1.82 \times 10^{-1}$\\
				& 0.30  &   $1.59 \times 10^{-1}$  &  0.30  &  $1.53 \times 10^{-1}$ & 0.30 &  $1.43 \times 10^{-1}$ \\
				& 0.60  &   $1.38 \times 10^{-1}$  &  0.60  &  $1.35 \times 10^{-1}$ & 0.60 &  $1.22 \times 10^{-1}$\\
				& 1.00    &   $1.15 \times 10^{-1}$  &  1.00    &  $1.14 \times 10^{-1}$ & 1.00 &  $1.03 \times 10^{-1}$\\
				1.0& 0.15 &   $1.25 \times 10^{+1} $  &  0.15 &  $1.21 \times 10^{+1}$& 0.15 &  $5.54 \times 10^{+0}$ \\
				&  0.30 &   $1.25 \times 10^{+1}$  &  0.30  &  $1.20 \times 10^{+1}$  & 0.30 &  $5.26 \times 10^{+0}$\\
				&  0.60 &   $1.41 \times 10^{+1}$  &  0.60  &  $1.36 \times 10^{+1}$ & 0.60 &  $5.63 \times 10^{+0}$\\
				&    1.00 &   $1.36 \times 10^{+1}$  &  1.00    &  $1.32 \times 10^{+1}$ & 1.00 &  $5.19 \times 10^{+0}$ \\
				1.5&0.15  &   $2.42 \times 10^{+2}$  &  0.15 &  $3.30 \times 10^{+2}$ & 0.15 &  $4.28 \times 10^{+1}$ \\
				&  0.30 &   $2.60 \times 10^{+2}$  &  0.30  &  $3.29 \times 10^{+2}$ & 0.30 &  $4.61 \times 10^{+1}$ \\
				&  0.60 &   $3.50 \times 10^{+2}$  &  0.60  &  $4.07 \times 10^{+2}$ & 0.60 &  $5.73 \times 10^{+1}$ \\
				&    1.00 &   $4.05 \times 10^{+2}$  &  1.00    &  $4.47 \times 10^{+2}$  & 1.00 &  $5.98 \times 10^{+1}$ \\
				\hline
	\end{tabular}}}
	
	\addtolength{\tabcolsep}{1pt}
	\scalebox{0.70}{
	{\begin{tabular}{ccccccc}
				\toprule
				\multicolumn{1}{c}{$^{92}$Zr(n,$\gamma$)$^{93}$Zr} 
				\\
				\cmidrule(rl){2-7}
				\multicolumn{1}{c}{} & \multicolumn{2}{c}{{BSFM}} & \multicolumn{2}{c}{{CTM}} &\multicolumn{2}{c}{{GSM}}\\
				\cmidrule(rl){2-3} \cmidrule(rl){4-5} \cmidrule(rl){6-7}
				$a$ & $k$T (MeV)& MACS (mb) & $k$T (MeV) & MACS (mb) & $k$T (MeV) & MACS (mb) \\
				\midrule
				0.5& 0.15 &   $1.14 \times 10^{-1}$  &  0.15 &  $4.64 \times 10^{-1}$& 0.15 &  $4.69 \times 10^{-1}$\\
				& 0.30  &   $7.79 \times 10^{-2}$  &  0.30  &  $3.34 \times 10^{-1}$ & 0.30 &  $3.34 \times 10^{-1}$ \\
				& 0.60  &   $4.74 \times 10^{-2}$  &  0.60  &  $2.10 \times 10^{-1}$ & 0.60 &  $2.09 \times 10^{-1}$\\
				& 1.00    &   $3.59 \times 10^{-2}$  &  1.00    &  $1.38 \times 10^{-1}$ & 1.00 &  $1.38 \times 10^{-1}$\\
				1.0& 0.15 &   $1.51 \times 10^{+1}$  &  0.15 &  $1.57 \times 10^{+1}$& 0.15 &  $2.16 \times 10^{+1}$ \\
				&  0.30 &   $1.30 \times 10^{+1}$  &  0.30  &  $1.36 \times 10^{+1}$  & 0.30 &  $1.91 \times 10^{+1}$\\
				&  0.60 &   $9.73 \times 10^{+0}$  &  0.60  &  $1.02 \times 10^{+1}$ & 0.60 &  $1.48 \times 10^{+1}$\\
				&    1.00 &   $6.89 \times 10^{+0}$  &  1.00    &  $7.38 \times 10^{+0}$ & 1.00 &  $1.09 \times 10^{+1}$ \\
				1.5&0.15  &   $3.33 \times 10^{+2}$  &  0.15 &  $3.54 \times 10^{+2}$& 0.15 &  $1.28 \times 10^{+2}$ \\
				&  0.30 &   $3.18 \times 10^{+2}$  &  0.30  &  $3.50 \times 10^{+2}$ & 0.30 &  $1.23 \times 10^{+2}$ \\
				&  0.60 &   $3.13 \times 10^{+2}$  &  0.60  &  $3.66 \times 10^{+2}$ & 0.60 &  $1.13 \times 10^{+2}$ \\
				&    1.00 &   $2.89 \times 10^{+2}$  &  1.00    &  $3.64 \times 10^{+2}$  & 1.00 &  $9.77 \times 10^{+1}$ \\
				\bottomrule
	\end{tabular}}}
\end{table}

{We have {also} computed the radiative capture rates of $n + {^{90,92}\rm{Zr}}\rightarrow {^{91,93}\rm{Zr_\gamma}}$ up to $T_9$=1 as mentioned in Table.~(\ref{tab:Rates}). {It was found that the  radiative capture rates of $n+{^{92}\rm{Zr}}\rightarrow{^{93}\rm{Zr_{\gamma}}}$ are higher by as much as 41\% compared to $n+{^{90}\rm{Zr}}\rightarrow{^{91}\rm{Zr_{\gamma}}}$  because of the stable nature of $^{90}\rm{Zr}$.} {{$^{90}\rm{Zr}$} has magic  neutron number ($N$=50), and therefore, has a smaller capture cross-section for incoming neutron than $^{92}\rm{Zr}$ ($N$=52). $^{92}\rm{Zr}$ does not offer any resistance {to} the incoming neutron. Therefore, the neutron capture rates of $^{92}\rm{Zr}$ are higher than $^{90}\rm{Zr}$. It is {apparent} that the rates determined in this work lead to a reduction in the $s$-component of $^{92}$Zr.}}

\begin{table}
	\centering
	\caption{Comparison of the radiative capture rates. The first column shows the temperature in the units of $10^{9}$~K, the second column shows the radiative capture rates of $n + {^{90}\rm{Zr}}\rightarrow {^{91}\rm{Zr_{\gamma}}}$, the third column show the radiative capture rates of  $n + {^{92}\rm{Zr}}\rightarrow {^{93}\rm{Zr_{\gamma}}}$, and the fourth column shows the percentage difference ($\%_{\Delta}$) between these rates.}
	\label{tab:Rates}       
	\addtolength{\tabcolsep}{1pt}
	\begin{tabular}{cccc}
		\toprule
		\multicolumn{1}{c}{} & \multicolumn{3}{c}{{Radiative capture rates ($cm^{3}mol^{-1}s^{-1}$)}}  \\
		\cmidrule(rl){2-4} 
		$T_9$ & $^{90}$Zr(n,$\gamma$)$^{91}$Zr& $^{92}$Zr(n,$\gamma$)$^{93}$Zr& $\%_{\Delta}$ \\
		\midrule
		0.0001 &$1.61\times10^{07}$&$2.26\times10^{07}$&{41.25} \\
		0.0005 &$1.04\times10^{07}$&$1.49\times10^{07}$&{41.35} \\
		0.001 &$8.35\times10^{06}$&$1.20\times10^{07}$&{43.71}  \\
		0.01 &$6.03\times10^{06}$&$8.40\times10^{06}$ &{39.30}  \\
		0.05 &$4.76\times10^{06}$&$6.76\times10^{06}$ &{42.02}  \\
		0.10 &$4.06\times10^{06}$&$5.81\times10^{06}$ &{43.10}  \\
		0.15 &$3.69\times10^{06}$&$5.27\times10^{06}$ &{42.82}  \\
		0.20 &$3.47\times10^{06}$&$4.95\times10^{06}$ &{42.65}  \\
		0.25 &$3.34\times10^{06}$&$4.74\times10^{06}$ &{41.92}  \\
		0.30 &$3.26\times10^{06}$&$4.60\times10^{06}$ &{41.10}  \\
		0.40 &$3.17\times10^{06}$&$4.43\times10^{06}$ &{39.75}  \\
		0.50 &$3.15\times10^{06}$&$4.35\times10^{06}$ &{38.10}  \\
%		0.60 &$3.16\times10^{06}$&$4.33\times10^{06}$ \\
%		0.70 &$3.19\times10^{06}$&$4.34\times10^{06}$ \\
%		0.80 &$3.23\times10^{06}$&$4.37\times10^{06}$ \\
%		0.90 &$3.30\times10^{06}$&$4.42\times10^{06}$ \\
%		01 &$3.36\times10^{06}$&$4.48\times10^{06}$ \\
		
		\bottomrule
	\end{tabular}
\end{table}

\newpage	
\section{Conclusion}
{In conclusion, we have computed the ($n$,$\gamma$) cross-section of the neutron magic and nearly magic number isotopes $^{90}$Zr and  $^{92}$Zr within the framework of the Talys v1.96 {code} over a wide range of energies. We deduced a lower capture rates for $^{90}$Zr(n,$\gamma$)$^{91}$Zr than $^{92}$Zr(n,$\gamma$)$^{93}$Zr, usually included in models of the $s$-process in AGB stars. It is to be noted that Talys gives a more or less similar representation of the data for the phenomenological sets of NLDs and GSFs. The MACS have been generated for the $^{90}$Zr(n,$\gamma$)$^{91}$Zr and $^{92}$Zr(n,$\gamma$)$^{93}$Zr processes via different NLD and GSF models found in Talys. {The generated values were then compared with the measurements in the literature}. Among them, we noted that the {{Kopecky-Uhl}} Lorentzian model of Talys is the best fit GSF model for $^{90}$Zr(n,$\gamma$)$^{91}$Zr and $^{92}$Zr(n,$\gamma$)$^{93}$Zr. The prediction of the Hauser-Feshbach theory, involving all the parameter adjustments and model variations, does successfully reproduce the available experimental data for both nuclei. Both of these processes are important for the formation of heavy elements in stars. The present investigations show that the rates of $^{92}$Zr(n,$\gamma$)$^{93}$Zr  process are higher than the radiative capture rates of $^{90}$Zr(n,$\gamma$)$^{91}$Zr, and confirms that $^{92}$Zr(n,$\gamma$)$^{93}$Zr has lower $s$-process component from the studied isotopes. }
\newpage

            %%%%%%%%%%%%%%%  BIBLIOGRAPHY  %%%%%%%%%%%%%% 

%\begin{table}[hbtp]
%	\centering
%	\caption{Effect of the adjustment parameter ($a$) on MACS for the $^{90}$Zr(n,$\gamma$)$^{91}$Zr and $^{92}$Zr(n,$\gamma$)$^{93}$Zr radiative capture processes. }
%	\label{tab:CTM1}       
%	\addtolength{\tabcolsep}{1pt}
%	\scalebox{0.70}{
%{\begin{tabular}{cccccccccc}
%\toprule
%\multicolumn{1}{c}{$^{90}$Zr(n,$\gamma$)$^{91}$Zr} 
%\\
%\cmidrule(rl){2-7}
%\multicolumn{1}{c}{} & \multicolumn{2}{c}{{R$_n$} (fm)} & \multicolumn{2}{c}{{R$_p$ (fm)}} &\multicolumn{2}{c}{{R$_{rms}$ (fm)}}\\
%\cmidrule(rl){2-3} \cmidrule(rl){4-5} \cmidrule(rl){6-7}
%Nuclei & L3R & This wr & X & y & Molar & This work \\
%\midrule
%$^{80}$Zr& 0.15 & 0.15  &  0.15 & 0.15 & 0.15 &  0.15\\
%$^{82}$Zr& 0.15 & 0.15  &  0.15 & 0.15 & 0.15 &  0.15\\
%$^{83}$Zr& 0.15 & 0.15  &  0.15 & 0.15 & 0.15 &  0.15\\
%$^{81}$Nb& 0.15 & 0.15  &  0.15 & 0.15 & 0.15 &  0.15\\
%$^{82}$Nb& 0.15 & 0.15  &  0.15 & 0.15 & 0.15 &  0.15\\
%$^{83}$Nb& 0.15 & 0.15  &  0.15 & 0.15 & 0.15 &  0.15\\
%			
%				\hline
%	\end{tabular}}}
%
%\end{table}

\end{document}